\documentclass[12pt]{article}
\usepackage{amsmath,amsfonts,amsthm,amssymb}
\usepackage{graphicx}

\usepackage{tikz}

\newcommand{\bra}[1]{\langle #1|}
\newcommand{\ket}[1]{|#1\rangle}

\newcommand{\Tr}{{\rm Tr}}

\newcommand{\eqnlabel}[1]{ \stepcounter{eqnnum} \tag{\arabic{eqnnum}} \label{#1} }
\newcounter{eqnnum}

\begin{document}
\setcounter{eqnnum}{0}

\begin{titlepage}
\hfill
\vbox{
    \halign{#\hfil         \cr
           } 
      }  
\vspace*{20mm}
\begin{center}
{\Large \bf Perturbative entanglement entropy\\in nonlocal theories}

\vspace*{15mm}
\vspace*{1mm}
Charles Rabideau
\vspace*{1cm}
\let\thefootnote\relax\footnote{rabideau@phas.ubc.ca}

{Department of Physics and Astronomy,
University of British Columbia\\
6224 Agricultural Road,
Vancouver, B.C., V6T 1W9, Canada}

\vspace*{1cm}
\end{center}

\begin{abstract}
Entanglement entropy in the vacuum state of local field theories exhibits an area law. However, nonlocal theories at large $N$ and strong coupling violate this area law. In these theories, the leading divergence in the entanglement entropy is extensive for regions smaller than the effective nonlocality scale and proportional to this effective nonlocality scale for regions larger than it.
This raises the question: is a volume law a generic feature of nonlocal theories, or is it only present at strong coupling and large $N$?

This paper investigates entanglement entropy of large regions in weakly coupled nonlocal theories, to leading order in the coupling. The two theories studied are $\phi^4$ theory on the noncommutative plane and $\phi^4$ theory with a dipole type nonlocal modification using a fixed nonlocality scale.
Both theories are found to follow an area law to first order in the coupling, hence no evidence is found for a volume law. This indicates that, perturbatively the nonlocal interactions considered are not generating sufficient entanglement at distances of the nonlocality scale to change the leading divergence, at least to first order in the coupling. An argument against volume laws at higher orders is also presented.
\end{abstract}
\end{titlepage}

\vskip 1cm


\section{Introduction and Summary}
Entanglement entropy has recently attracted interest as a way to study the correlations between degrees of freedom in a quantum state.
Local field theories generally exhibit what is know as an area law behaviour, where the leading divergence in the entanglement entropy of a spatial region 
is proportional to the area of the boundary of that region. That is, 
$S \sim |\partial A| \Lambda^{d-2}$,
where $S$ is the entanglement entropy, $|\partial A|$ the area of the boundary of the region and $\Lambda$ is the momentum scale of the UV regulator of the theory, for example the inverse of a lattice spacing.\footnote{See for example \cite{Eisert:2008ur} for a review of area laws in entanglement entropy.}
However, recent holographic studies of strongly coupled nonlocal theories have found a volume law behaviour instead \cite{Barbon:2008ut, Li:2010dr, Fischler:2013gsa, Karczmarek:2013xxa,Pang:2014tpa}. That is, for a nonlocality scale $l$, $S \sim |A| \Lambda^{d-1}$ for regions much smaller than $l$ and $S \sim l |\partial A| \Lambda^{d-1}$ for regions much larger than $l$ \cite{Karczmarek:2013xxa}. 
Note that entanglement entropy of large regions is sufficient to differentiate this type of volume law from an area law, as the entanglement entropy is proportional to the length scale of the nonlocality times an additional factor of the UV regulator. To summarise,
\begin{align*}
\mathrm{area~law:}&   &S &\sim |\partial A| \Lambda^{d-2}, && \\
\mathrm{volume~law:}&   & S &\sim |A| \Lambda^{d-1}, &(\mathrm{small~regions})& \\
 &  &S &\sim l |\partial A| \Lambda^{d-1}. &(\mathrm{large~regions})&
\end{align*}

These results can be understood intuitively by assuming that all the degrees of freedom within the range of the nonlocality are equally entangled with each other. Then, for regions much smaller than $l$, all the degrees of freedom inside the region, not only those near the boundary, are entangled with degrees of freedom outside. For regions much larger than $l$, all the degrees of freedom within a distance $l$ of the boundary are entangled with those outside. In both cases, the number of degrees of freedom strongly entangled across the boundary is proportional to $\Lambda^{d-1}$ rather than the $\Lambda^{d-2}$ expected from an area law.

A natural question is whether this behaviour is generic to nonlocal theories or if it is confined to a strongly coupled, large $N$ regime. One approach is to study entanglement entropy for a free scalar field on the fuzzy sphere \cite{Dou:2006ni,Dou:2009cw,Karczmarek:2013jca,Sabella-Garnier:2014fda}. This turns out to be proportional to the area\footnote{The fuzzy sphere is a 2 dimensional surface, thus $|A|$ is an area and $|\partial A|$ is a circumference.} for small polar caps \cite{Karczmarek:2013jca,Sabella-Garnier:2014fda}. However, two issues arise which question whether this should be characterised as a volume law. First, the dependence of the entanglement entropy on the UV regulator does not match the volume law described above. Second, the entanglement entropy does not scale like the number of degrees of freedom contained in the polar cap, as the degrees of freedom are not uniformly distributed across the sphere. Instead it scales as the number of degrees of freedom near the boundary \cite{Dou:2006ni,Dou:2009cw}. Another limitation of this theory is that the nonlocality scale is tied to the size of the sphere so it is not possible to study regions much larger than the nonlocality scale. 

Another approach is to study a free field theory on a lattice with a nonlocal kinetic term, in which case a volume law was found \cite{Shiba:2013jja}.

This paper investigates the role of interactions in this question by considering two theories with nonlocal interactions: scalar $\lambda \phi^4$ theory on the noncommutative plane and $\lambda \phi^4$ theory with a dipole type nonlocal modification with fixed nonlocality scale.
The leading divergence in entanglement entropy of large regions is calculated to leading order in perturbation theory and is not found to be proportional to the length scale of the nonlocality, hence no evidence of a volume law is found. Instead, the leading divergence in both theories has the same form as the standard local $\lambda \phi^4$ theory which follows an area law. This result indicates that, perturbatively these nonlocal interactions are not generating sufficient entanglement at distances of the nonlocality scale to change the leading divergence, at least to first order in the coupling.

The free theory with $\lambda=0$ for both of these nonlocal theories is equivalent to the regular commutative $\lambda\phi^4$ theory. There is no modification of the entanglement entropy at this order. Perturbation theory can be used to study the nonlocal theories at small $\lambda$.

The entanglement entropy is calculated using the replica trick and the formula $S = -\partial_n \left[ \ln Z_n - n \ln Z_1 \right]_{n=1}$, where $Z_n$ is the partition function of the field theory defined on an $n$-sheeted space \cite{Calabrese:2004eu,Casini:2009sr,Hertzberg:2012mn}. This partition function can be reduced to computing vacuum bubble diagrams and the $O(\lambda)$ contribution in perturbation theory comes from bubble diagrams with one vertex and two loops. Consistent with the results of previous investigations of perturbative noncommutative theories \cite{Minwalla:1999px}, the planar diagrams in the nonlocal theories give the standard commutative result, which is $S \sim G_1(0) \int dx \partial_{n=1} G_n(x) \sim A_\perp \Lambda^2 \ln (\Lambda/m)$, where $A_\perp$ is the (infinite) area of the boundary of our region, $\Lambda$ our UV regulator, $m$ our IR regulator and $G_n$ is the Green's function on the $n$-sheeted space used in the replica trick \cite{Hertzberg:2012mn}. This contribution follows an area law, as $S \propto A_\perp \Lambda^2$ up to logarithmic corrections.

The nonlocality only affects the nonplanar diagram. This diagram contributes a term of the form $S \sim G_1(0,\Delta x) \int dx \partial_{n=1} G_n(x,x+\Delta x) \sim \frac{A_\perp}{(\Delta x)^2} \ln f(\Lambda,m,\Delta x)$, where now $\Delta x$ corresponds to a translation from the nonlocality. 

In the dipole theory, $\Delta x$ is proportional to the fixed dipole length. Thus the nonplanar diagram has only a logarithmic IR divergence and is subleading compared to the planar diagrams. 
In the noncommutative theory the translation along the noncommuative plane is proportional to the momentum in the other noncommutative direction, so this contribution must be integrated over this momentum.  
If we don't impose an IR regulator, the momentum controlling the translation is allowed to vanish and $G(0,\Delta x) \rightarrow G_1(0) \sim \Lambda^2$. This gives a contribution that is of the same order as the planar diagrams. 
However, if we impose an IR regulator, $\Delta x$ has a minimal value and this divergence can be reinterpreted as an IR divergence. This is familiar from the UV/IR connection described for example in \cite{Minwalla:1999px}. 

Our results for the $O(\lambda)$ contribution to the entanglement entropy, $S_1$, are
\begin{align*}
&\textrm{real scalar}:& S_{1} =& 2 \lambda S_\textrm{planar} 
+ \lambda S_\textrm{nonplanar}\\
&\textrm{charged scalar}:& S_{1} =& (2\lambda_0 + \lambda_1) S_\textrm{planar} 
+ \lambda_1 S_\textrm{nonplanar},
\end{align*}
where $S_\textrm{planar}$ and $S_\textrm{nonplanar}$ denote the contributions from planar and nonplanar diagrams respectively.

The leading divergences from these diagrams in each of the theories considered are
\begin{align*}
&&S_\textrm{planar} 
=& - \frac{ A_\perp \Lambda^2}{2^{10} 3^2 \pi^3} 
\ln \frac{\Lambda^2}{4 m^2}\\
&\textrm{Commutative theory}:& S_\textrm{nonplanar} =& S_\textrm{planar} \\
&\textrm{Noncommutative plane}:& S_\textrm{nonplanar}
=& - \frac{A_\perp \Lambda^2}{2^{9} 3^2 \pi^3} 
\frac{-\ln\left(\frac{\Theta^2 m^2 \Lambda^2}{4}\right)}
{1 - \frac{\Theta^2 m^2 \Lambda^2}{4}} + \mathrm{subleading}\\
&\textrm{Dipole theory}:& S_\textrm{nonplanar} \mathrm{~is~} & \mathrm{subleading} ,
\end{align*}
where $\Lambda$ is our UV regulator, $m$ is our IR regulator, $A_\perp$ is the area of the boundary, $\Theta$ is the noncommutativity parameter of the plane and $a$ is the nonlocality scale of the dipole theory. The details of the expansion in $\frac{m}{\Lambda}$ used to extract these leading divergences are discussed in Section \ref{sec:expansion}.

In both cases, the contribution from these nonplanar diagrams does not have the right form to be interpreted as the sign of a volume law in the entanglement entropy and we must conclude that these nonlocal theories at least to first order in perturbation theory obey an area law.
This can be contrasted with the strong coupling result which found clear signs of the volume law even for large regions \cite{Karczmarek:2013xxa}. Thus, the volume law must either only appear at higher orders in perturbation theory or it must require strong coupling. Consistent with our analysis, previous investigations of perturbative dynamics of the noncommutative theory \cite{Minwalla:1999px} have shown that noncommutativity does not introduce any new perturbative UV divergences that cannot be reinterpreted as IR divergences. Thus, is it hard to see how the higher degree of divergence required for a volume law can arise in perturbation theory. We are lead to the conclusion that entanglement on distances of the nonlocality scale and volume laws require strong coupling and are not accessible to perturbation theory.


The remainder of the paper is organised as follows: Section \ref{sec:theories} describes the theories we study, Section \ref{sec:EE} explains how the entanglement entropy can be computed perturbatively in these theories, Section \ref{sec:free} shows that the results for the free theory are unchanged in these nonlocal theories, Section \ref{sec:comm} computes the first order correction in the coupling to the entanglement entropy in a real scalar $\phi^4$ theory for a warm-up and for later reference. Section \ref{sec:NC} extends the calculation to the real scalar on the noncommutative plane. Section \ref{sec:charged} reproduces the results for the previous two sections in the case of the charged scalar. Section \ref{sec:dip} computes the result for the charged scalar in the dipole theory. Finally, Section \ref{sec:conclusion} concludes with a discussion of these results.

\section{Theories}
\label{sec:theories}
The theories used in this paper are scalar field theories on $\mathbb{R}^{1,3}$ where products of fields are replaced with a possibly noncommutative product denoted $\star$. Three examples of this product will be used: the regular commutative one, the Moyal product associated with the noncommutative plane and the dipole product with a fixed nonlocality scale. 
See \cite{Douglas:2001ba} for a review of noncommutative field theory.
The Euclidean action is
\begin{align*}
\mathcal{S}_E = \int d^d x \left[ -\frac12 \partial \phi \star \partial \phi (x) 
+ \frac12 m^2 \phi \star \phi (x)
+ \frac{\lambda}{4!} \phi \star \phi \star \phi \star \phi (x) \right].
\end{align*}
The entanglement entropy in these three theories is calculated to leading order in the coupling $\lambda$. The mass is present to serve as an IR regulator and will be taken to be small in the end.

First, the standard commutative case, where $(f \star g) (x) = f(x) g(x)$, is reviewed and presented in our notation in Sections \ref{sec:free} through \ref{sec:comm}. The entanglement entropy for this theory was studied in \cite{Hertzberg:2012mn} and the approach contained therein will be followed for each of the theories we consider.

Second, in Section \ref{sec:NC}, the entanglement entropy of a field theory defined on the noncommutative plane, where
\begin{align*}
(f \star g) (x) = \exp\left(\frac{i}{2} \Theta^{\mu\nu} \frac{\partial}{\partial \xi^\mu} \frac\partial{\partial \zeta^\nu} \right) f(x+\xi) g(x+\zeta) |_{\xi=\zeta=0},
\end{align*}
is studied. The noncommutativity is parametrised by the antisymmetric tensor $\Theta$. This theory has been studied perturbatively in \cite{Minwalla:1999px}. In this case especially, the mass should be thought of as an IR regulator and taken to zero at the end of the calculation in order to see full effects of the UV/IR mixing present in this theory. We specialise to the case commonly referred to as the noncommutative plane where $\Theta^{\mu\nu} = \Theta \left( \delta^{1\mu}\delta^{2\nu} - \delta^{2\mu}\delta^{1\nu} \right)$ for simplicity.

Finally, the entanglement entropy of the a simpler nonlocal theory with a fixed nonlocality scale along a particular axis, known as a dipole theory, where
\begin{align*}
(f \star g) (x) = f(x + \frac12 Q_L(g)) 
g(x - \frac12 Q_L(f))
\end{align*}
is studied.

This requires a global charge to identify with the dipole length. This can be obtained by using a charged scalar rather than the real scalar field theory we have discussed so far. The action for a complex scalar is
\begin{align*}
\mathcal{S}_E = \int d^d x \left[ -\partial \phi^\dagger \star \partial \phi (x) 
+ m^2 \phi^\dagger \star \phi (x)
+ \frac{\lambda_0}{4} \phi^\dagger \star \phi \star \phi^\dagger \star \phi (x) 
+ \frac{\lambda_1}{4} \phi^\dagger \star \phi \star \phi \star \phi^\dagger (x) \right].
\end{align*}
where there two $\phi^4$ terms which are inequivalent due to our noncommutative product \cite{Dasgupta:2001zu}.\footnote{These noncommutative products are constructed to ensure that integrals of products of fields are invariant under cyclic permutations.} 

The result from the real scalar theory will be extended to this charged scalar theory in Section \ref{sec:charged}, then the dipole theory will be studied in Section \ref{sec:dip}. 

Setting $Q_L = \vec a Q_{U(1)}$, where $\vec a= a \hat \imath$ is the fixed dipole length and $\hat\imath$ is a unit vector in the $x_1$ direction, the terms in the action can be written in a more explicit form:
\begin{align*}
\int dx (\phi^\dagger \star \phi)(x) 
&= \int dx \phi^\dagger(x + \frac12 a) \phi(x + \frac12 a)
= \int dx \phi^\dagger(x) \phi(x), \\
\int dx (\phi^\dagger \star \phi) \star (\phi^\dagger \star \phi) (x)
&= \int dx \phi^\dagger(x) \phi(x) \phi^\dagger(x) \phi(x), 
\eqnlabel{eqn:dip_identities}\\
\int dx (\phi^\dagger \star \phi) \star (\phi \star \phi^\dagger) (x)
&= \int dx \phi^\dagger(x + \frac12 a) \phi(x+ \frac12 a) \phi(x - \frac12 a) \phi^\dagger(x - \frac12 a). 
\end{align*}

In fact, renormalisability requires that we include terms $\lambda_n \int dx (\phi^\dagger \phi)(x+ \frac12 n a) (\phi^\dagger \phi)(x - \frac12 n a)$ for all $n$ \cite{Dasgupta:2001zu}. However, the contributions from these terms can be obtained by simply substituting $a \rightarrow n a$ into the results for $n=1$ and summing over $n$. The results in Section \ref{sec:dip} are such that this sum is guaranteed to converge as long as the $\lambda_n$ don't grow too quickly. As the inclusion of these terms would not affect our conclusions, we will not consider them separately. 

\section{Entanglement Entropy}
\label{sec:EE}
The standard technique of the replica trick is used to compute the entanglement entropy \cite{Calabrese:2004eu}. This technique was used in a perturbative context in \cite{Hertzberg:2012mn}, whose approach is followed here.

Starting with $\rho_A$, the reduced density matrix of the ground state of the theory in question for a region $A$, the idea is to evaluate
\begin{align*}
S = -\Tr(\rho_A \ln \rho_A) = - \frac{\partial}{\partial n} \ln \Tr(\rho_A^n) |_{n=1},
\eqnlabel{eqn:def_EE}
\end{align*}
by calculating $\Tr \rho_A^n$ for arbitrary $n$ and analytically continuing. In this paper we will concentrate on the simplest case where $A$ is the half plane ($A = \{(x_1,x_2,x_3) \in \mathbb{R}^3 | x_1>0 \}$).

The main result that will be needed can be lifted directly from \cite{Calabrese:2004eu,Hertzberg:2012mn}:
\begin{align*}
\ln \Tr(\rho_A^n) = \ln Z_n - n \ln Z_1, 
\eqnlabel{eqn:def_Tr_p_n}
\end{align*}
where $Z_n$ is the partition function of the theory on an $n$-sheeted surface with a cut along the region $A$ that connects the sheets.
However, some details of this $n$-sheeted space will be needed in the argument to follow, so the rest of this section will define it more carefully.

\subsection{\textit{n}-sheeted surfaces}
The density matrix can be written as a path integral, (at finite inverse temperature of $\beta$)
\begin{align*}
\bra{\phi_2} \rho \ket{\phi_1} 
= \left( Z_1 \right)^{-1} 
\int \mathcal{D}\phi_{\phi(x,0) = \phi_1}^{\phi(x,\beta) = \phi_2}
e^{-\mathcal{S}_E},
\end{align*}
where $Z_1$ is a normalisation factor to ensure that $\Tr \rho = 1$.
Then the reduced density matrix for a region $A$ is obtained by periodically identifying the field in the Euclidean time direction along $\bar A$, the complement of $A$, while leaving the  boundary condition along $A$ untouched. 
To look at the ground state, $\beta$ must be sent to infinity. We do this while keeping the cut along $A$ near the origin.

Then,
\begin{align*}
\Tr (\rho_A^n) = \left( Z_1 \right)^{-n} 
\left[ \int \mathcal{D}\phi_{\phi(x\in A,0^+) = \phi_1}^{\phi(x\in A,0^-) = \phi_2} 
e^{-\mathcal{S}_E} \right] 
\left[ \int \mathcal{D}\phi_{\phi(x\in A,0^+) = \phi_2}^{\phi(x\in A,0^-) = \phi_3} 
e^{-\mathcal{S}_E} \right]
 \ldots 
\left[ \int \mathcal{D}\phi_{\phi(x\in A,0^+) = \phi_n}^{\phi(x\in A,0^-) = \phi_1} 
e^{-\mathcal{S}_E} \right].
\end{align*}
This identification of boundary conditions can be replaced by defining the field theory on an $n$-sheeted surface with a cut along $A$ that takes you from one sheet to the next. Calling this $n$-sheeted surface $\left( \mathbb{R}^d \setminus A \right)^n$, the projection onto the sheet $\pi: \left( \mathbb{R}^d \setminus A \right)^n \rightarrow \mathbb{R}^d \setminus A$ and the indicator function telling you if you are on the $k^{\mathrm{th}}$ sheet $\chi_k: \left( \mathbb{R}^d \setminus A \right)^n \rightarrow \mathbb{Z}_1$,
this means that $\Phi: \left( \mathbb{R}^d \setminus A \right)^n \rightarrow \mathbb R$ can be defined as $\Phi(x) = \sum_{k=1}^N \phi_k(\pi(x)) \chi_k(x)$, so that
\begin{align*}
\Tr (\rho_A^n) = \left( Z_1 \right)^{-n} 
\left[ \int \mathcal{D}\Phi e^{-\mathcal{S}_E} \right],
\end{align*}
where $\mathcal{S}_E$ for $\Phi$ has the same form as that for each $\phi$, since the action for each sheet is additive.

With our simple region $A$, a half-plane, polar coordinates can be defined in the $x$-$\tau$ plane of $\mathbb{R}^d \setminus A$. Then the glueing required to create this $n$-sheeted surface is simply to identify $\theta = 2\pi$ on one sheet to $\theta = 0$ on the next. Thus polar coordinates can be defined on $\left( \mathbb{R}^d \setminus A \right)^n$ where $\theta \in [0,2\pi n)$, such that each interval of length $2 \pi$ corresponds to a sheet, \textit{i.e.} $\pi(r,\theta,y,z) = (r,\theta \textrm{ mod } 2\pi,y,z)$ and $\chi_k(r,\theta,y,z) = \chi_{[2 \pi (k-1),2 \pi k)}(\theta)$. 
\label{sec:polar coordinates}

This gives us the result from \cite{Calabrese:2004eu,Hertzberg:2012mn} cited above, as $ Z_n =  \int \mathcal{D}\Phi e^{-\mathcal{S}_E}$. This path integral over $\Phi$ is the path integral over the $n$-sheeted surface.

\section{Free Theory}
\label{sec:free}
The first step is to understand the free theories where $\lambda=0$. 
The action for the free noncommutative and dipole theories is the same for that of the commutative theory, since the star product of 2 fields is the same as the regular product up to a total derivative \cite{Minwalla:1999px}. 

For the noncommutative theory,
\begin{align*}
\int d^4 x (f \star g) (x) =& \int d^4 x~ \sum_{n=0}^\infty ~ \frac{i^n}{2^n} ~
\Theta^{\mu_1 \nu_1} \ldots \Theta^{\mu_n \nu_n}~ 
\partial_{\mu_1} \ldots \partial_{\mu_n} f(x) ~
\partial_{\nu_1} \ldots \partial_{\nu_n} g(x)
\\
=& \int d^4 x \left[ f(x) g(x) 
+ \partial_{\mu_1} ~ \sum_{n=1}^\infty ~
\Theta^{\mu_1 \nu_1} \ldots \Theta^{\mu_n \nu_n} ~
\partial_{\mu_2} \ldots \partial_{\mu_n} f(x) ~
\partial_{\nu_1} \ldots \partial_{\nu_n} g(x) \right],
\end{align*}
so that the quadratic term in the action is the same as for the commutative case up to a total derivative. As there are no boundaries, the only place this total derivative could make for a finite contribution is at the conical singularity introduced at the origin when considering the $n$-sheeted path integral.

Around the origin this term contributes (note that the singularity is at the origin of the $x$-$\tau$ plane and is not localised in the $y$-$z$ directions),
\begin{align*}
\lim_{r \rightarrow 0} A_\perp \sum_{n=1}^\infty \int r d \theta ~ \Theta^{r\nu_1}
 \Theta^{\mu_2 \nu_2} \ldots \Theta^{\mu_n \nu_n} ~
 \partial_{\mu_2} \ldots \partial_{\mu_n} \phi ~\partial_{\nu_1} \ldots \partial_{\nu_n} \phi
  \sim  \lim_{r \rightarrow 0} \sum_n r \partial^n\phi\partial^{n+1}\phi,
\end{align*}
where $A_\perp$ is the area of the $y$-$z$ plane. As long as $\partial^n\phi~\partial^{n+1}\phi$ is regular at the origin this term will not contribute to the action. This means that $\phi$ needs to be $C^\infty$ at the origin, which is just the regular boundary condition imposed in the commutative case.

For the dipole theory, direct calculation of the $\star$-product of two fields can be seen to reduce to the commutative result in Equation \eqref{eqn:dip_identities}. 

Thus the free theory is the same for all three theories.

\subsection{Green's functions}
Since the free theories are the same, they have the same Green's functions. This Green's function is straightforward in the polar coordinates introduced in Section \ref{sec:polar coordinates}. Since the action for $\Phi$ living on the $n$-sheeted surface is the same as the action for $\phi$ living on any particular sheet, the local equation that the Green's function must obey will be the same. The only difference is that $\theta$ must be periodic with period $2\pi n$ rather than the usual period of $2\pi$. The Green's function for the field living on the $n$-sheeted surface is, from \cite{Hertzberg:2012mn},
\begin{align*}
G_n(x,x') = \frac1{2\pi n}\int \frac{d^{d_\perp} p^\perp}{(2 \pi)^{d_\perp}} \sum_{k=0}^\infty a_k 
\int_0^\infty dq q \frac{J_{k/n}(qr) J_{k/n}(qr')}{q^2+p^2_\perp+m^2} \cos(k (\theta-\theta')/n) 
e^{i p_\perp (x_\perp - x'_\perp)},
\eqnlabel{eqn:def_G_n}
\end{align*}
where $a_0 = 1$, $a_{k\neq0} = 2$, $p_\perp = (p_y,p_z)$ and $x_\perp = (x_2,x_3)$. $\perp$ refers to the directions orthogonal to the cut introduced by the replica trick.

The Euler-Maclaurin formula,
\begin{align*}
\sum_{k=0}^\infty a_k  F(k) = 2 \left[ \int_0^\infty dk F(k) \right] - \frac16 F'(0) - 2 \sum_{j>1} \frac{B_{2j}}{(2j)!} F^{(2j-1)}(0),
\end{align*}
can be applied to this Green's function to replace the sum over $k$,
\begin{align*}
G_n(x,x') = &\int_0^\infty \frac{dk}{\pi}\int \frac{d^{d_\perp} p^\perp}{(2 \pi)^{d_\perp}} 
\int_0^\infty dq q \frac{J_{k}(qr) J_{k}(qr')}{q^2+p^2_\perp+m^2} \cos(k (\theta-\theta')) 
e^{i p_\perp (x_\perp - x'_\perp)} \\
&-\frac1{12\pi n^2}\int \frac{d^{d_\perp} p^\perp}{(2 \pi)^{d_\perp}} 
\int_0^\infty dq q \frac{\partial_\nu [J_{\nu}(qr) J_{\nu}(qr')]_{\nu=0}}{q^2+p^2_\perp+m^2} 
e^{i p_\perp (x_\perp - x'_\perp)}  \eqnlabel{eqn:def_j_terms}\\
&-\sum_{j>1} \frac{B_{2j}}{\pi n^{2j} (2j)!}\int \frac{d^{d_\perp} p^\perp}{(2 \pi)^{d_\perp}} 
\int_0^\infty dq q \frac{(\partial_\nu)^{2j-1} 
[J_{\nu}(qr) J_{\nu}(qr') \cos(\nu(\theta-\theta'))]_{\nu=0}}{q^2+p^2_\perp+m^2} 
e^{i p_\perp (x_\perp - x'_\perp)}. 
\end{align*}

It will be useful to define $G_n(x,x';p)$ as
\begin{align*}
G_n(x,x';p_y) = \frac1{2\pi n} \int \frac{d p_z}{2 \pi} \sum_{k=0}^\infty a_k
\int_0^\infty dq q \frac{J_{k/n}(qr) J_{k/n}(qr')}{q^2+p_y^2 + p_z^2 +m^2} 
\cos(k (\theta-\theta')/n) 
e^{i p_z (x_3 - x'_3) + i p_y (x_2-x'_2)} \eqnlabel{eqn:def_G_n_3_params}
\end{align*}
such that
\begin{align*}
G_n(x,x') &= \int \frac{dp_y}{2 \pi} G_n(x,x';p_y) \\
\frac\partial{\partial x_2} G_n(x,x';p) &= - \frac\partial{\partial x'_2} G_n(x,x';p) = i p G_n(x,x';p).
\end{align*}

It is also useful to define $f_n(x,x')$ and $f_n(x,x';p)$ as
\begin{align*}
f_n(x,x') &= G_n(x,x') - G_1(x,x') \eqnlabel{eqn:def_f} \\
&= \frac{n^2-1}{12\pi n^2}\int \frac{d^{d_\perp} p^\perp}{(2 \pi)^{d_\perp}} 
\int_0^\infty dq q \frac{\partial_\nu [J_{\nu}(qr) J_{\nu}(qr')]_{\nu=0}}{q^2+p^2_\perp+m^2} 
e^{i p_\perp (x_\perp - x'_\perp)} + (j>1)\\
f_n(x,x';p) &= G_n(x,x';p) - G_1(x,x';p),
\end{align*}
where $G_1$ is the Green's function on the $1$-sheeted surface, that is just the regular Green's function.

\subsubsection{Single sheeted limit}
This Green's function for the n-sheeted space must reduce to the regular Green's function in the limit where $n \rightarrow 1$. Starting with our expression for the Green's function in Equation \eqref{eqn:def_G_n}, defining $\varphi = \theta-\theta'$ for convenience and setting $n = 1$,
\begin{align*}
G_1(x,x') = \frac1{2\pi}\int \frac{d^{d_\perp} p^\perp}{(2 \pi)^{d_\perp}} \sum_{k=0}^\infty a_k 
\int_0^\infty dq q \frac{J_{k}(qr) J_{k}(qr')}{q^2+p^2_\perp+m^2} \cos(k \varphi) 
e^{i p_\perp (x_\perp - x'_\perp)}.
\end{align*}

Equation (10.9.E2) in the DLMF \cite{NIST:DLMF} provides a useful integral representation of the Bessel functions, which can be rewritten as,
$
J_n (z) =  \int_{-\pi}^\pi \frac{d\gamma}{2\pi} e^{i(z\sin\gamma - n \gamma)}.
$
Using this representation and the fact that $J_{-k}(z) = (-1)^k J_k(z)$, \footnote{Equation (10.4.E1) in \cite{NIST:DLMF}}
\begin{align*}
\sum_{k=0}^\infty a_k J_{k}(qr) J_{k}(qr') \cos(k \varphi)
=& \sum_{k=-\infty}^\infty \int_{-\pi}^\pi  \frac{d\gamma d\kappa}{(2\pi)^2}
e^{iq(r\sin\gamma + r'\sin\kappa) - i k (\gamma+\kappa)} e^{ik\varphi} \\
=& \int_{-\pi}^\pi \frac{d\gamma}{2\pi} e^{iq\left[r\sin\gamma + r'\sin(\varphi-\gamma) \right]}.
\end{align*}

Defining our position axes on the $x_0$-$x_1$ plane such that $\vec{x} = (0,r)$ implies that 
$\vec{x}' = (-r'\sin\varphi, r'\cos\varphi)$. Then defining $\vec{q} = (q\cos\gamma, q\sin\gamma)$,
\begin{align*}
\vec{q}\cdot (\vec{x} - \vec{x}') =& q \left[r\sin\gamma + r' \sin(\varphi-\gamma)\right] \\
\sum_{k=0}^\infty a_k J_{k}(qr) J_{k}(qr') \cos(k \varphi)
=& \int_{-\pi}^\pi \frac{d\gamma}{2\pi} e^{i \vec{q}\cdot (\vec{x} - \vec{x}')}
\end{align*}
Finally, defining $p = (\vec{q},p_\perp)$,
\begin{align*}
G_1(x,x') =  \int \frac{d^{d} p}{(2 \pi)^{d}} 
\frac{e^{i p (x - x')}}{p^2+m^2},
\end{align*}
which is the usual Euclidean Green's function.

\subsection{Entanglement entropy in the free theory}
The entanglement entropy when $\lambda=0$ must be identical in the three theories as it was shown above that the quadratic terms in the action are the same. This can be seen more explicitly by using the approach from \cite{Hertzberg:2012mn}.
Starting from $S_A = -\partial_n \left[ \ln Z_{n} - n \ln Z_{1}\right]_{n=1}$, the part of the entanglement entropy which depends on the mass can be related to the Green's function by
\begin{align*}
\frac{\partial}{\partial m^2} \ln Z_{n} = -\frac12 \int_n d^d x \langle \Phi^2(x) \rangle_n.
\end{align*}

In the commutative case, $\langle \Phi^2(x) \rangle_n = G_n(x,x)$. In the non-commutative case,
\begin{align*}
\langle \Phi \star \Phi (x) \rangle_n 
=& \left(\exp\left[\frac{i}{2} \Theta 
\left(\frac\partial{\partial \xi_1} \frac{\partial}{\partial \zeta_2} 
-\frac\partial{\partial \xi_2} \frac\partial{\partial \zeta_1} \right)\right]
\langle \Phi(x+\xi) \Phi(x+\zeta) \rangle_n \right)_{\xi=\zeta=0} \\
=& \left(\exp\left[\frac{i}{2} \Theta 
\left(\frac\partial{\partial \xi_1} \frac{\partial}{\partial \zeta_2} 
-\frac\partial{\partial \xi_2} \frac\partial{\partial \zeta_1} \right)\right]
G_n(x+\xi,x+\zeta) \right)_{\xi=\zeta=0} \\
=& \int \frac{dp_y}{2\pi} \left(\exp\left[\frac{1}{2} \Theta p_y 
\left(\frac\partial{\partial \xi_1}
+\frac\partial{\partial \zeta_1} \right) \right]
G_n(x+\xi,x+\zeta;p_y) \right)_{\xi=\zeta=0} \\
=& \int \frac{dp_y}{2\pi} 
G_n(x + \frac{1}{2} \Theta p_y \hat{\imath},x + \frac{1}{2} \Theta p_y \hat{\imath};p_y).
\end{align*}
 That the $\star$-product turns out to just translate the argument of the Green's function is an important theme of the calculation in this paper.

The only difference for a complex scalar is that the mass term in the action is proportional to $\Phi^\dagger \star \Phi$ instead of $\Phi \star \Phi$, however the expectation value of this leads to the same Green's function and the same result follows. 

The dipole theory is identical except that translations by $\Theta$ times the momentum in the $y$-direction are replaced by translations by $a$.

Thus, still for the non-commutative case, 
\begin{align*}
\frac{\partial}{\partial m^2} \ln Z_{n} 
=& -\frac12 \int_n d^d x \langle \Phi \star \Phi (x) \rangle_n\\
=& -\frac12 \int_n d^d x \int \frac{dp}{2\pi} 
G_n(x + \frac{1}{2} \Theta p \hat{\imath},x + \frac{1}{2} \Theta p \hat{\imath};p) \\
=& -\frac12 \int_n d^d x \int \frac{dp}{2\pi} 
G_n(x,x;p) = -\frac12 \int_n d^d x G_n(x,x),
\end{align*}
recovering explicitly  the result from the commutative case by shifting the integration variable. 

\begin{figure}
  \centering
    \includegraphics[width=.80\textwidth]{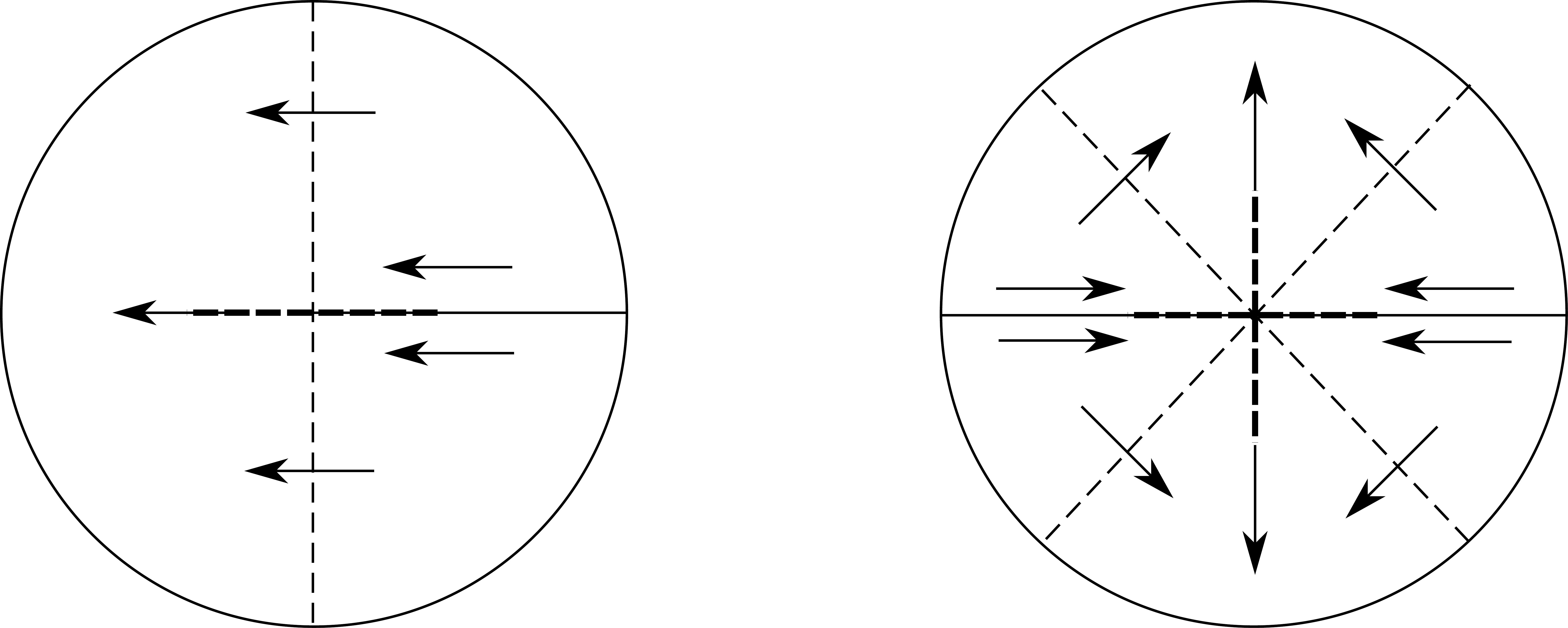}\
  \caption[Translations on the $n$-sheeted surface.]{Translations on each of the sheets of the $n$-sheeted surface (on the left) give a well defined map on the whole surface (shown for $n=2$ in the polar coordinates described in Section \ref{sec:polar coordinates} on the right), except for a measure zero set near the singularity at the origin.}
  \label{fig:shift_cone}
\end{figure}

However, this shift of the integration variable on the $n$-sheeted surface bears further investigation. It is sketched in Figure \ref{fig:shift_cone}.

This shift is well defined except for the region which gets translated into or out of the origin. However, this region has measure zero and cannot affect the result of the integral. As long as only a countable number of such shifts are done, these points can  be omitted from the integral without changing the result. Finally, the integral over the whole $n$-sheeted surface can be written as a sum over the sheets and the Jacobian of this shift on each sheet is $1$, so the Jacobian of the whole shift does not introduce any new factors into the integral. Thus shifting the variable of integration on this $n$-sheeted surface is allowed with no Jacobian, just as for the plane.

\section{First Order in Perturbation Theory}
\subsection{Commutative theory}
\label{sec:comm}
We will start by computing the first order correction to the entanglement entropy for the commutative $\phi^4$ theory. This was done previously in \cite{Hertzberg:2012mn}, but will be repeated here with more explicit regulators that will allow a direct comparison to the nonlocal cases.
From \cite{Hertzberg:2012mn},
\begin{align*}
\ln Z_n &= \ln \int \mathcal{D}\phi e^{-S_E[\phi]}\\
&= \ln Z_{n,0} - \frac{\lambda}{4!} \int_n d^4 x 
\langle \Phi^4 (x) \rangle_0 + ...\\
&=  \ln Z_{n,0} - \frac{3\lambda}{4!} \int_n d^4 x 
\left[ G_n(x,x) \right]^2 + ..., \eqnlabel{eqn:comm_wick}
\end{align*}
where $\int_n$ denotes integration over the $n$-sheeted surface and $\ln Z_{n,k}$ is the $k^\textrm{th}$ order term in a $\lambda$ expansion of $\ln Z_n$. Generally, adding subscript will denote the order of a term in a $\lambda$ expansion, e.g. $X = X_0 + X_1 + X_2 + \ldots$

The entanglement entropy can be calculated using Equations \eqref{eqn:def_EE} and \eqref{eqn:def_Tr_p_n},
\begin{align*}
\ln \Tr \left( \rho_A^n \right)_1 =& \ln Z_{n,1} - n \ln Z_{1,1} \\
=& - \frac{3\lambda}{4!} \int_n d^4 x 
\left[ G_n(x,x) \right]^2 + \frac{3 n \lambda}{4!} \int d^4 x 
\left[ G_1(x,x) \right]^2 \\
=& - \frac{3\lambda}{4!} \int_n d^4 x 
\left[ 2 G_1(x,x) f_n(x,x) + f_n^2(x,x) \right].
\end{align*}

Recalling from Equation \eqref{eqn:def_f}, 
\begin{align*}
f_n(x,x') = \frac{n^2-1}{12\pi n^2}\int \frac{d^{d_\perp} p^\perp}{(2 \pi)^{d_\perp}} 
\int_0^\infty dq q \frac{\partial_\nu [J_{\nu}(qr) J_{\nu}(qr')]_{\nu=0}}{q^2+p^2_\perp+m^2} 
e^{i p_\perp (x_\perp - x'_\perp)} + (j>1).
\end{align*}
The $j>1$ terms don't contribute \cite{Calabrese:2004eu}, so they will be dropped in what follows. 
This is the same on each sheet, so the integral over the $n$-sheeted surface is $n$ times in integral on one sheet. Finally, $f_1(x,x')=0$, so $\partial_n f_n^2(x,x') |_{n=1} = 0$ and
\begin{align*}
S_1 = - \partial_n \left[ \ln \Tr \left( \rho_A^n \right)_1 \right]_{n=1} 
=\frac{6\lambda}{4!} \int d^4 x  G_1(x,x) \partial_n \left[ n f_n(x,x) \right]_{n=1} \eqnlabel{eqn:S1_comm}
\end{align*}

\begin{align*}
S_1 = \frac{12 \lambda A_\perp}{12 \pi \cdot 4!} \int r dr d\phi \int \frac{d^4 k d p_y d p_z}{(2\pi)^6} \frac{1}{k^2 +m^2} 
\int_0^\infty dq q \frac{\partial_\nu [J_{\nu}(qr) J_{\nu}(qr)]_{\nu=0}}{q^2+p_y^2 + p_z^2 + m^2}.
\end{align*} 

Schwinger parameters are introduced to allow the denominators to be combined, using 
\begin{align*}
\frac{1}{A} = \int_0^\infty d\alpha e^{-A\alpha}.
\end{align*}
This allows us to regulate the UV divergence in $S_1$ by introducing a factor of $e^{-\frac{1}{\alpha \Lambda^2}}$, as was done in previous perturbative studies of noncommutative theories \cite{Minwalla:1999px}. This regulator is convenient in the noncommutative case and is used here so that the results can be compared. Using Equation (25) from p.146 in volume I of \cite{ET},
\begin{align*}
\int_0^\infty dt e^{-pt -\frac{a}{4t}} = \sqrt{\frac{a}{p}} K_1(\sqrt{a p}),
\eqnlabel{eqn:ET_146-25}
\end{align*}
the effect of this regulator is
\begin{align*}
\int_0^\infty d\alpha e^{-\alpha p^2 - \frac{1}{\alpha \Lambda^2}} 
= \frac{2}{\Lambda p} K_1\left(\frac{2p}{\Lambda}\right) &\xrightarrow{\frac{p}{\Lambda} \rightarrow \infty} \sqrt{\frac{2}{\Lambda p^3}} e^{-\frac{2p}{\Lambda}},\\
&\xrightarrow{\frac{p}{\Lambda} \rightarrow 0} \frac{1}{p^2}.
\end{align*}
Thus it regulates the UV and leaves the IR unaffected. This can be seen simply from the fact that $e^{-\frac{1}{\alpha \Lambda^2}}$ vanishes for $\alpha \ll \Lambda^{-2}$ and goes to one for $\alpha \gg \Lambda^{-2}$. A mass $m$ regulates the IR by contributing a factor of $e^{-\alpha m^2}$, which has the opposite behaviour.

Introducing these Schwinger parameters and regulating,
\begin{align*}
S_1 = \frac{\lambda A_\perp}{3 \cdot 2^3} \int dr \frac{d^4 k d p_y d p_z}{(2\pi)^6} dq 
\int_0^\infty d \alpha d \beta 
qr e^{-\alpha k^2 - \beta \left[q^2+p_y^2 + p_z^2 \right] -\alpha m^2 - \frac{1}{\alpha \Lambda^2} - \beta m^2 - \frac{1}{\beta \Lambda^2} }
\partial_\nu [J_{\nu}(qr) J_{\nu}(qr)]_{\nu=0}.
\end{align*} 
All the momenta integrals except $q$ are Gaussian,
\begin{align*}
S_1 =& \frac{\lambda A_\perp}{3 \cdot 2^{9} \pi^3} \int dr dq 
\int_0^\infty d \alpha d \beta 
\frac{qr}{\alpha^2 \beta} e^{- \beta q^2 
-\alpha m^2 - \frac{1}{\alpha \Lambda^2} - \beta m^2 - \frac{1}{\beta \Lambda^2} }
\partial_\nu [J_{\nu}(qr) J_{\nu}(qr)]_{\nu=0}.
\end{align*} 

Using Equation (10.22.E67) from the Digital Library of Mathematical Functions (DLMF) \cite{NIST:DLMF},
\begin{align*}
\int_0^\infty t e^{-p^2 t^2} J_\nu(at) J_\nu(bt) dt = \frac1{2p^2} e^{-\frac{(a^2+b^2)}{4p^2}} I_\nu \left(\frac{ab}{2p^2} \right),
\eqnlabel{eqn:DLMF_10.22.E67}
\end{align*}
the $q$ integral can be evaluated. This along with the fact that $\partial_\nu I_\nu(z) |_{\nu=0} = - K_0(z)$\footnote{Equation (10.38.E4) in the DLMF \cite{NIST:DLMF}.}
gives
\begin{align*}
S_1 =& -\frac{\lambda A_\perp}{3 \cdot 2^{10}\pi^3} \int dr 
\int_0^\infty d \alpha d \beta \frac{r}{\alpha^2 \beta^2} 
e^{ -\frac{r^2}{2\beta}
-\alpha m^2 - \frac{1}{\alpha \Lambda^2} - \beta m^2 - \frac{1}{\beta \Lambda^2} }
K_0 \left( \frac{r^2}{2\beta} \right).
\end{align*}

Equation (21) on p. 131 of \cite{ET},
\begin{align*}
\int_0^\infty d t e^{-a t} K_0(t y) 
= \frac{\arccos(\frac{a}{y})}{\sqrt{y^2 - a^2}} 
\xrightarrow{\frac{a}{y} \rightarrow 1} \frac{1}{y},
\eqnlabel{eqn:ET_131-21}
\end{align*}
after substituting $r^2 \rightarrow t$ and setting $a=y=\frac{1}{2\beta}$, gives
\begin{align*}
S_1 =& -\frac{\lambda A_\perp}{3 \cdot 2^{10} \pi^3} 
\left( \int_0^\infty \frac{d \alpha}{\alpha^2} 
e^{-\alpha m^2 - \frac{1}{\alpha \Lambda^2}} \right)
\left( \int_0^\infty \frac{d \beta}{\beta} 
e^{- \beta m^2 - \frac{1}{\beta \Lambda^2} } \right).
\end{align*} 

Looking at the $\alpha$ integral first,
\begin{align*}
\int_0^\infty \frac{d \alpha}{\alpha^2} 
e^{-\alpha m^2 - \frac{1}{\alpha \Lambda^2}}
=& \int_0^\infty d \alpha
e^{-\frac{ m^2 }{\alpha} - \frac{\alpha}{\Lambda^2}} \\
=& 2 m \Lambda K_1 \left( \frac{2m}{\Lambda} \right) \xrightarrow{\frac{m}{\Lambda} \rightarrow 0} \Lambda^2
\end{align*}
by substituting $\alpha \rightarrow \frac{1}{\alpha}$ in the first line and using Equation \eqref{eqn:ET_146-25} as well as in the second. This recovers the $\Lambda^2$ divergence seen previously in this case \cite{Hertzberg:2012mn}.

Using Equation (29) from Volume 1, p. 146 of \cite{ET} 
\begin{align*}
\int_0^\infty t^{\nu-1} e^{-pt -\frac{a}{4t}} dt
=& 2 \left(\frac{a}{4p}\right)^{\frac{\nu}{2}} K_\nu(\sqrt{ap})
\end{align*}
the $\beta$ integral gives,
\begin{align*}
\int_0^\infty \frac{d\beta}{\beta} e^{- \beta m^2 - \frac{1}{\beta \Lambda^2} }
=& 2 K_0 \left( \frac{2m}{\Lambda} \right) \xrightarrow{\frac{m}{\Lambda} \rightarrow 0}
-2\ln \frac{2m}{\Lambda} = \ln \frac{\Lambda^2}{4 m^2},
\end{align*}
as $K_0(z) \rightarrow -\ln z$ as $z \rightarrow 0$. This reproduced the logarithmic divergence seen previously in this case \cite{Hertzberg:2012mn} and makes explicit its form in our regularisation scheme.

Combining, the first order in $\lambda$ correction to the entanglement entropy in the commutative theory is
\begin{align*}
S_{1,\textrm{Comm.}} =& -3 \lambda \frac{A_\perp \Lambda^2}{3^2 \cdot 2^{10} \pi^3} 
\ln \frac{\Lambda^2}{4 m^2}.
\eqnlabel{eqn:dS_comm_result}
\end{align*}
This is proportional to the area of the boundary of $A$, that is $A_\perp$, and the leading divergence is of order $\Lambda^2$, so this result fits with the area law picture discussed in the introduction.

\subsection{Noncommutative theory}
\label{sec:NC}
Next we will compute the first order correction to the entanglement entropy for the noncommutative $\phi^4$ theory.
Similarly to the commutative theory,
\begin{align*}
\ln Z_n &= \ln \int \mathcal{D}\phi e^{-S_E[\phi]}\\
&= \ln Z_{n,0} - \frac{\lambda}{4!} \int_n d^4 x 
\langle \Phi \star \Phi \star \Phi \star \Phi (x) \rangle_0 + ...
\end{align*}

Using the associativity of the $\star$-product, this can be written as
\begin{align*}
\int_n d^4 x \langle \Phi \star \Phi \star \Phi \star \Phi (x) \rangle_0 
= \int_n d^4 x   &\left( 
\exp\left[\frac{i}{2} \Theta 
\left(\frac\partial{\partial \xi_1} \frac{\partial}{\partial \zeta_2} 
-\frac\partial{\partial \xi_2} \frac\partial{\partial \zeta_1} \right)\right] 
\right)_{\xi=\zeta=0}\\
&\left( 
\exp\left[\frac{i}{2} \Theta 
\left(\frac\partial{\partial \eta_1} \frac{\partial}{\partial \varsigma_2} 
-\frac\partial{\partial \eta_2} \frac\partial{\partial \varsigma_1} \right)
\right]
\right)_{\eta=\varsigma=0} \\
&\left( 
\exp\left[\frac{i}{2} \Theta 
\left(\frac\partial{\partial \gamma_1} \frac{\partial}{\partial \kappa_2} 
-\frac\partial{\partial \gamma_2} \frac\partial{\partial \kappa_1} \right)
\right]
\right)_{\gamma=\kappa=0}\\
&\qquad\qquad \langle \Phi(x+\xi+\eta)\Phi(x+\xi+\varsigma)\Phi(x+\zeta+\gamma)\Phi(x+\zeta+\kappa) \rangle.
\end{align*}

The usual Wick's Theorem can be applied to calculate the four-point function,
\begin{align*}
\langle \Phi(w)\Phi(x)\Phi(y)\Phi(z) \rangle = G_n(w,x) G_n(y,z) 
+ G_n(w,y) G_n(x,z) + G_n(w,z) G_n(x,y). \eqnlabel{eqn:NC_wick}
\end{align*}

The key point is that while the conical singularity breaks the translational invariance in the $x_0$-$x_1$ plane, it is preserved in the $x_2$-direction. Thus the star product reduces to a translation in the $x_1$-direction by an amount determined by the momentum in the $x_2$-direction. Defining $G_n(w,z) = \int \frac{d p_y}{2 \pi} G_n(w,z;p_y)$ as in Equation \eqref{eqn:def_G_n_3_params},
\begin{align*}
\exp\left(\frac{i}{2} \Theta \frac\partial{\partial {w}_1} \frac\partial{\partial {z}_2} \right) G_n(w,z)
&= \int \frac{d p_y}{2 \pi} \exp\left(\frac{1}{2} p_y \Theta \frac\partial{\partial {w}_1} \right)  G_n(w,z;p_y)  \\
&= \int \frac{d p_y}{2 \pi} G_n(w + \frac12 p_y \Theta \hat \imath,z;p_y),
\end{align*}
this can be used to evaluate the 4-point function,
\begin{align*}
\int_n d^4 x < \Phi \star \Phi &\star \Phi \star \Phi (x) >_0  \\
= &\int_n d^4 x   
\int \frac{d k_y d p_y }{(2\pi)^2} 
\bigg[ G_n(x+\frac12  \Theta k_y \hat\imath,x+\frac12 \Theta k_y \hat\imath;k_y) G_n(x+\frac12 \Theta p_y \hat\imath,x+\frac12 \Theta p_y \hat\imath;p_y)  \\
&+ G_n(x+\frac12 \Theta k_y \hat\imath,x+\frac12 \Theta (k_y+2p_y) \hat\imath;k_y) G_n(x+\frac12 \Theta (2k_y+p_y) \hat\imath,x+\frac12 \Theta p_y \hat\imath;p_y) \\
&+ G_n(x+\frac12 \Theta k_y \hat\imath,x+\frac12 \Theta k_y \hat\imath;k_y) G_n(x+\frac12 \Theta (2k_y+p_y) \hat\imath,x+\frac12 \Theta (2k_y+p_y) \hat\imath;p_y) \bigg] .
\end{align*}

Then, by shifting the spatial integral, 
\begin{align*}
= &\int_n d^4 x   
\int \frac{d k_y d p_y }{(2\pi)^2} 
\bigg[ G_n(x,x;k_y) G_n(x+\frac12 \Theta (p_y-k_y) \hat\imath,x+\frac12 \Theta (p_y-k_y) \hat\imath;p_y)  \\
&+ G_n(x-\frac12 \Theta p_y \hat\imath,x+\frac12 \Theta p_y \hat\imath;k_y) G_n(x+\frac12 \Theta k_y \hat\imath,x-\frac12 \Theta k_y \hat\imath;p_y) \\
&+ G_n(x,x;k_y) G_n(x+\frac12 \Theta (k_y+p_y) \hat\imath,x+\frac12 \Theta (k_y+p_y) \hat\imath;p_y) \bigg] .
\end{align*}

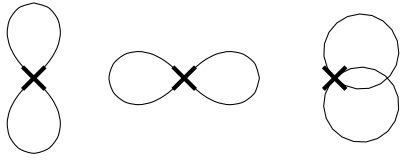
\begin{figure}
  \centering
\begin{tikzpicture}
\draw [ultra thick] (-.15,-.15) -- (.15,.15);
\draw [ultra thick] (-.15,.15) -- (.15,-.15);
\draw [ultra thick,shift={(2,0)}] (-.15,-.15) -- (.15,.15);
\draw [ultra thick,shift={(2,0)}] (-.15,.15) -- (.15,-.15);
\draw [ultra thick,shift={(4,0)}] (-.15,-.15) -- (.15,.15);
\draw [ultra thick,shift={(4,0)}] (-.15,.15) -- (.15,-.15);
\draw [domain=0:2*pi,smooth,samples=31] plot ( {sin(\x r)*cos(\x r)/(1+sin(\x r)*sin(\x r))} , {cos(\x r)/(1+sin(\x r)*sin(\x r))} );
\draw [domain=0:2*pi,smooth,samples=31] plot ({2+cos(\x r)/(1+sin(\x r)*sin(\x r))} , {sin(\x r)*cos(\x r)/(1+sin(\x r)*sin(\x r))} );
\begin{scope}[shift={(4,0)}]
\draw [domain=0:pi,samples=20,rotate=-45] plot ( {sin(\x r)*cos(\x r)},{sin(\x r) *sin(\x r)});
\draw [domain=0:pi,samples=20,rotate=-45] plot ( {cos(\x r)*cos(\x r)},{cos(\x r) *sin(\x r)});
\end{scope}
\end{tikzpicture} 
  \caption[Non-planar diagrams and non-commutivity.]{Vacuum bubble diagrams at leading order in a real scalar $\lambda \phi^4$ theory. The only vacuum bubble where lines cross is the second one. This is the only one which is affected by the non-commutativity, as discussed in \cite{Minwalla:1999px}.}
  \label{fig:non-planar}
\end{figure}

In \cite{Minwalla:1999px} it is seen that the effects of the non-commutativity manifest themselves in the diagrams where lines cross each other. This is also present here, as Figure \ref{fig:non-planar} shows that it is only the second term that involves lines crossing. The other two terms are two self-coincident Green's functions -- the same result as was found in the commutative case in Section \ref{sec:comm} and \cite{Hertzberg:2012mn}. The second term, which corresponds to the nonplanar diagram, is the only one which is different than what was found in the commutative case.

The entanglement entropy can be calculated using Equation \eqref{eqn:def_EE},
\begin{align*}
S_{1} =& -  \partial_n \left[ \ln Z_{n,1} - n \ln Z_{1,1} \right]_{n=1} 
\eqnlabel{eqn:EE_NC_start}\\
=& \frac{2 \lambda}{4!} \partial_n \Bigg( \int d^4 x   
\bigg[ 2 G_1(x,x) n f_n(x,x)  
+ \int \frac{d k_y d p_y }{(2\pi)^2}  G_1(x,x+\Theta p_y \hat\imath;k_y) 
n f_n(x,x-\Theta k_y \hat\imath;p_y) \bigg] \Bigg)_{n=1} 
\end{align*}
where the fact that the spatial integral can be shifted, that the momenta can be renamed, that $G_1(x,x;p_y) = G_1(x+a,x+a;p_y)$, that $f_n(x,x',p_y)=f_n(x,x';-p_y)$ as long as $x_2 = x'_2$ and that $f_1 = 0$ so that the terms with $f_n^2$ can be ignored have all been used. The $j>1$ terms in $f_n$ have also been dropped again, which allows us here to write the integral over the $n$-sheeted surface as $n$ times the integral over a sheet. In the commutative case, it was clear that these $j>1$ terms do not contribute \cite{Calabrese:2004eu}. In Appendix \ref{sec:appendix_j_terms} it is argued that the leading divergence must be entirely contained in the $j=1$ term even in this noncommutative theory.

\subsubsection{New contribution from the nonplanar diagram}
The first term in Equation \eqref{eqn:EE_NC_start} is the contribution from the two planar diagrams. These give the same result as in the commutative case, namely $\frac{\lambda A_\perp \Lambda^2}{2^{10} 3^2 \pi^3} \ln \frac{\Lambda^2}{4 m^2}$ from each diagram. 
However, the nonplanar diagram gives a new contribution to the entanglement entropy from the non-commutativity. The contribution from this nonplanar diagram will be denoted $S_\textrm{nonplanar}$,
\begin{align*}
S_\textrm{nonplanar} =& \frac{2 \lambda}{4!} \int d^4 x   
\int \frac{d k_y d p_y }{(2\pi)^2}  G_1(x,x+\Theta p_y \hat\imath;k_y) 
\partial_n \left[ n f_n(x,x-\Theta k_y \hat\imath;p_y) \right]_{n=1} \\
=& \frac{4 \lambda A_\perp}{12 \pi \cdot 4!} \int r dr d\phi \int \frac{d^4 k d p_y d p_z}{(2\pi)^6} \frac{e^{i \Theta k_x p_y}}{k^2+m^2} 
\int_0^\infty dq q \frac{\partial_\nu [J_{\nu}(qr) J_{\nu}(qr')]_{\nu=0}}{q^2+p_y^2 + p_z^2 + m^2},
\end{align*} 
where $r'^2 = (\vec{r} - \Theta k_y \hat\imath)^2 = r^2 + (\Theta k_y)^2 - 2 \Theta r k_y \cos\phi$ and $A_\perp$ is the area of the $x_2$-$x_3$ plane that bounds the region for which the entanglement entropy is being calculated.

The next step is to introduce Schwinger parameters and to regulate this integral in the same manner as the integrals for other perturbative calculations in this noncommutative theory were regulated in \cite{Minwalla:1999px}, as discussed in Section \ref{sec:comm},
\begin{align*}
S_\textrm{nonplanar} = \frac{\lambda A_\perp}{2^3 3^2 \pi} \int dr d\phi \frac{d^4 k d p_y d p_z}{(2\pi)^6} dq 
\int_0^\infty d \alpha d \beta 
qr e^{-\alpha k^2 - \beta \left[q^2+p_y^2 + p_z^2 \right] 
- \frac{1}{\alpha \Lambda^2} -\alpha m^2 
- \frac{1}{\beta \Lambda^2} -\beta m^2 }  \\
e^{i \Theta k_x p_y} \partial_\nu [J_{\nu}(qr) J_{\nu}(qr')]_{\nu=0}.
\end{align*}
 
The $p_y$, $p_z$ and $k$ except for $k_y$ integrals are all Gaussian (recall that $r'$ is a function of $k_y$),
\begin{align*}
S_\textrm{nonplanar} =& \frac{\lambda A_\perp }{2^{8} 3^2 \pi^{\frac92}} \int dr d\phi dk_y dq 
\int_0^\infty d\alpha d\beta 
\frac{qr}{\alpha\sqrt\beta \sqrt{4 \alpha \beta + \Theta^2}} \\
&\qquad e^{-\alpha k_y^2 - \beta q^2 
- \frac{1}{\alpha \Lambda^2} -\alpha m^2 - \frac{1}{\beta \Lambda^2} -\beta m^2 }
\partial_\nu [J_{\nu}(qr) J_{\nu}(qr')]_{\nu=0}.
\end{align*} 

In order to make explicit some of the symmetry between $r$ and $r'$, $\rho$ and $\varphi$ can be defined such that $r = \rho \sin\varphi$ and $k_y = \frac{\rho}{\Theta} \cos\varphi$, with $\rho \in [0,\infty)$ and $\varphi \in [0,\pi]$. Then defining $g(\phi,\varphi) = \sqrt{1 + \sin2\varphi \cos\phi}$, gives $r' = \rho g(\phi,\varphi)$ in these variables. Performing this change of variables,
\begin{align*}
S_\textrm{nonplanar} =& \frac{\lambda A_\perp }{2^{8} 3^2 \pi^{\frac92} \Theta} 
\partial_\nu|_{\nu=0} \int d\rho d\varphi d\phi dq d\alpha d\beta 
\frac{q \rho^2 \sin\varphi }{\alpha\sqrt\beta \sqrt{4 \alpha \beta + \Theta^2}} \\
&\qquad e^{-\frac\alpha{\Theta^2} \rho^2 \cos^2\varphi - \beta q^2 
- \frac{1}{\alpha \Lambda^2} -\alpha m^2 - \frac{1}{\beta \Lambda^2} -\beta m^2 } 
J_{\nu}(q\rho \sin\varphi) J_{\nu}(q\rho g(\phi,\varphi)).
\end{align*} 

From the DLMF (10.22.E67) \cite{NIST:DLMF}, 
\begin{align*}
\int_0^\infty t e^{-p^2 t^2} J_\nu(at) J_\nu(bt) = \frac1{2p^2} e^{-\frac{(a^2+b^2)}{4p^2}} I_\nu \left(\frac{ab}{2p^2} \right)
\eqnlabel{eqn_NC:DLMF_10.22.E67}
\end{align*}
so that,
\begin{align*}
S_\textrm{nonplanar} =& \frac{\lambda A_\perp }{2^{9} 3^2 \pi^{\frac92} \Theta} 
\partial_\nu|_{\nu=0} \int d\rho d\varphi d\phi d\alpha d\beta 
\frac{\rho^2 \sin\varphi }{\alpha \beta^{\frac32} \sqrt{4 \alpha \beta + \Theta^2}} \\
&e^{-\frac\alpha{\Theta^2} \rho^2 \cos^2\varphi 
-\rho^2 \frac{\sin^2\varphi + g^2(\varphi,\phi)}{4 \beta}
- \frac{1}{\alpha \Lambda^2} -\alpha m^2 - \frac{1}{\beta \Lambda^2} -\beta m^2 } 
I_{\nu}\left(\frac{\rho^2}{2\beta} g(\phi,\varphi) \sin\varphi \right).
\end{align*} 

Now $\rho$ and $\alpha$ can be rescaled to simplify this expression as $\rho \rightarrow 2 \sqrt{\beta} \rho$ and $\alpha \rightarrow \frac{\Theta^2}{4\beta} \alpha$,
\begin{align*}
S_\textrm{nonplanar} =& \frac{\lambda A_\perp }{2^6 3^2 \pi^{\frac92} \Theta^2} 
\partial_\nu|_{\nu=0} \int d\rho d\varphi d\phi d\alpha 
\frac{\rho^2 \sin\varphi}{\alpha \sqrt{\alpha + 1}}  
\left(\int_0^\infty d\beta 
e^{- \beta \left( m^2 + \frac{4}{\Theta^2 \Lambda^2 \alpha} \right) 
- \frac{1}{\beta} \left( \frac{1}{\Lambda^2} + \frac{\Theta^2 m^2 \alpha}{4} \right) } \right)
\\
&e^{- \alpha \rho^2 \cos^2\varphi -\rho^2 \left[\sin^2\varphi + g^2(\phi,\varphi)\right]} I_{\nu}\left(2 \rho^2 g(\phi,\varphi) \sin\varphi \right).
\end{align*} 

Equation (25) from p.146 in volume I of \cite{ET},
\begin{align*}
\int_0^\infty dt e^{-pt -\frac{a}{4t}} = \sqrt{\frac{a}{p}} K_1(\sqrt{a p}),
\eqnlabel{eqn_NC:ET_146-25}
\end{align*}
allows the $\beta$ integral to be evaluated,
\begin{align*}
\int_0^\infty d\beta 
e^{- \beta \left( m^2 + \frac{4}{\Theta^2 \Lambda^2 \alpha} \right) 
- \frac{1}{\beta} \left( \frac{1}{\Lambda^2} + \frac{\Theta^2 m^2 \alpha}{4} \right) } 
=& \sqrt{\frac{\frac{4}{\Lambda^2} + \Theta^2 m^2 \alpha}
{m^2 + \frac{4}{\Theta^2 \Lambda^2 \alpha}}}
K_1\left( \sqrt{ \left(\frac{4}{\Lambda^2} + \Theta^2 m^2 \alpha\right)
\left(m^2 + \frac{4}{\Theta^2 \Lambda^2 \alpha} \right)} \right), \\
=& \Theta \sqrt\alpha 
K_1\left( \frac{4}{\Theta \Lambda^2 \sqrt\alpha} + \Theta m^2 \sqrt\alpha \right).
\end{align*}

Using the identity $\partial_\nu|_{\nu=0} I_\nu(z) = -K_0(z)$,
\begin{align*}
S_\textrm{nonplanar} =& - \frac{\lambda A_\perp }{2^6 3^2 \pi^{\frac92} \Theta} 
\int d\rho d\varphi d\phi d\alpha 
\frac{\rho^2 \sin\varphi}{\sqrt\alpha \sqrt{ \alpha + 1}}  
e^{-\rho^2 \left[ \alpha \cos^2\varphi + \sin^2\varphi + g^2(\phi,\varphi)\right]} \\
&K_0\left(2 \rho^2 g(\phi,\varphi) \sin\varphi \right) 
K_1\left(\frac{4}{\Theta \Lambda^2 \sqrt\alpha} + \Theta m^2 \sqrt\alpha \right).
\end{align*} 

Taking a large $\Lambda$ limit of this expression and expanding $K_1(x) \approx \frac1x$ for $x \rightarrow 0$ allows us to extract an overall quadratic divergence. However, more progress can still be made by evaluating the $\rho$ integral.

Using in order Equation (23) from p. 131 of \cite{ET} and (15.9.E19) of \cite{NIST:DLMF},
\begin{align*}
\int_0^\infty d\rho \rho^2 e^{-A \rho^2} K_0(B \rho^2)
=& \int_0^\infty dx \sqrt{x} e^{-A x} K_0(B x) \\
=& \frac12 \sqrt{\pi} \frac{[\Gamma(\frac32)]^2}{\Gamma(2) (A+B)^{\frac32}} 
{}_2F_1 \left(\frac32,\frac12;2;\frac{A-B}{A+B}\right) \\
=& \frac{\pi^{\frac32}}{8\sqrt2 B^{\frac32}} \frac{1}{\sqrt{\left(\frac{A}{B}\right)^2 - 1} } 
P^1_{-\frac12}\left(\frac{A}{B}\right),
\end{align*}
where $P^1_{-\frac12}(x)$ is the appropriate branch of the associated Legendre function with non-integer degree.

Defining $z =  \frac{\alpha \cos^2\varphi + \sin^2\varphi + g^2(\varphi,\phi)}{2 g(\phi,\varphi) \sin\varphi }$ and recalling that $g(\phi,\varphi) = \sqrt{1 + \sin2\varphi \cos\phi}$,
\begin{align*}
S_\textrm{nonplanar}
=& - \frac{\lambda A_\perp}{2^{11} 3^2 \pi^3 \Theta} 
\int_0^\infty d\alpha \frac{G(\alpha)}{\sqrt\alpha \sqrt{ \alpha + 1}}
K_1\left(\frac{4}{\Theta \Lambda^2 \sqrt\alpha} + \Theta m^2 \sqrt\alpha \right) \mathrm{~and}\\
G(\alpha) =& \int_0^\pi d\varphi \int_0^{2\pi} d\phi
\frac{1}{\left[ g(\phi,\varphi) \right]^{\frac32} \sqrt{\sin\varphi} }
\frac{P^1_{-\frac12}(z)}{\sqrt{z^2-1}},
\end{align*}
where $G(\alpha)$ is dimensionless and finite for $\alpha \in (0,\infty)$. 

At this point, the asymptotic behaviour of $G(\alpha)$ can be analysed numerically, as no analytic formula for this integral was found in the tables consulted. However, while analysing this asymptotic behaviour, we found that $G(\alpha) = \frac{16}{\sqrt{\alpha + 1}}$ gives an exact match up to high numerical accuracy across the many orders of magnitude that were checked.\footnote{The only potential divergences in the integral for $S_\textrm{nonplanar}$ come from the regions of small and large $\alpha$. If the reader is uncomfortable with this numeric argument, this functional form for $G(\alpha)$ could also be thought of more conservatively as a function with the right asymptotic behaviour to reproduce the correct divergences in this integral.}

Using this result for $G(\alpha)$,
\begin{align*}
S_\textrm{nonplanar} =& - \frac{\lambda A_\perp}{2^{7} 3^2 \pi^3 \Theta} 
\int_0^\infty \frac{d\alpha}{\sqrt\alpha} \frac{1}{\alpha + 1}
K_1\left(\frac{4}{\Theta \Lambda^2 \sqrt\alpha} + \Theta m^2 \sqrt\alpha \right).
\end{align*}
Note that this result is invariant under $\Theta \Lambda^2 \leftrightarrow \Theta m^2$, another sign of the UV/IR connection in non-commutative theories.

\label{sec:expansion}
This integral has two regulators, $\Lambda$ and $m$. The only other dimensionful parameter is $\Theta$, so the only dimensionless products of these regulators are $\frac{m}{\Lambda}$ and $\Theta m \Lambda$. As is familiar from the UV/IR mixing in this theory, the limits $\Lambda \rightarrow \infty$ and $m \rightarrow 0$ do not commute. This can be resolved by taking $\frac{m}{\Lambda} \rightarrow 0$ while fixing $\Theta m \Lambda$. Then taking the limit $m \rightarrow 0$ or $\Lambda \rightarrow \infty$ first corresponds to the limits $\Theta m \Lambda \rightarrow 0$ or $\Theta m \Lambda \rightarrow \infty$ respectively.  
\footnote{This discussion applies even if we want to think of $m$ as a physical mass, as the ratio $\frac{m}{\Lambda}$ will still vanish if $m$ is fixed while $\Lambda \rightarrow \infty$. This case corresponds to $\Theta m \Lambda \rightarrow \infty$.}

Introducing $\gamma = \sqrt{\alpha}$,
\begin{align*}
S_\textrm{nonplanar} =& - \frac{\lambda A_\perp}{2^{6} 3^2 \pi^3 \Theta} 
\int_0^\infty d\gamma \frac{1}{\gamma^2 + 1}
K_1\left(\frac{2m}{\Lambda} \left[\frac{2}{\Theta m \Lambda \gamma} 
+ \frac{\Theta m \Lambda \gamma}{2} \right]\right) \\
\xrightarrow{\frac{m}{\Lambda} \rightarrow 0}&
- \frac{\lambda A_\perp \Lambda}{2^{8} 3^2 \pi^3 \Theta m} 
\int_0^\infty d\gamma \frac{1}{\gamma^2 + 1}
\frac{1}{\frac{2}{\Theta m \Lambda \gamma} 
+ \frac{\Theta m \Lambda \gamma}{2}} \\
=& - \frac{\lambda A_\perp \Lambda^2}{2^{9} 3^2 \pi^3} 
\frac{-\ln\left(\frac{\Theta^2 m^2 \Lambda^2}{4}\right)}
{1 - \frac{\Theta^2 m^2 \Lambda^2}{4}} 
= - \frac{\lambda A_\perp }{2^{7} 3^2 \pi^3 \Theta^2 m^2} 
\frac{-\ln\left(\frac{4}{\Theta^2 m^2 \Lambda^2}\right)}
{1-\frac{4}{\Theta^2 m^2 \Lambda^2}},
\eqnlabel{eqn:dS_NCYM_result}
\end{align*}
where the last line uses Equation (2) from Volume 2 p.216 of \cite{ET}.

This result illustrates the UV/IR connection in non-commutative theories. 
If the IR regulator is removed first ($\Theta m \Lambda \ll 1$), $S_\textrm{nonplanar} \sim A_\perp \Lambda^2$ -- a quadratic UV divergence. However if the UV regulator is removed first ($\Theta m \Lambda \gg 1)$, $S_\textrm{nonplanar} \sim \frac{A_\perp}{\Theta^2 m^2} $, allowing the same divergence to be interpreted as an IR divergence. 
In addition, whether $\frac{\Theta^2 m^2 \Lambda^2}{4}$ is taken to be large or small there is a logarithmic divergence as is found in the commutative case. 
However, here there is the additional option of keeping both regulators, that is keeping $\frac12 \Theta m \Lambda$ finite, which eliminates the logarithmic divergence seen in the commutative case.\footnote{Note that if a $\Theta \rightarrow 0$ limit is taken, this option is no longer available and the commutative result is recovered, although the exact form of the logarithmic divergence depends on how the $\Theta$ limit is taken.} 
In particular, there is a natural choice of IR regulator\footnote{See Section 6 of \cite{Minwalla:1999px}}, $m = \frac{2}{\Theta \Lambda}$ where
\begin{align*}
S_\textrm{nonplanar} =& - \frac{\lambda A_\perp }{2^{7} 3^2 \pi^3 \Theta^2 m^2}
 = - \frac{\lambda A_\perp \Lambda^2}{2^{9} 3^2 \pi^3}.
\end{align*}

From a mathematical point of view, this UV/IR connection can be seen to originate from the translation of the arguments of the Green's function. In the commutative theory, $S_\textrm{nonplanar} \sim \int_n dx G_n(x,x) f_n(x,x)$ where as in the noncommutative theory, the non-planar diagram made a contribution of the form $S_\textrm{nonplanar} \sim \int_n dx G_n(x,x+\Theta p) f_n(x,x+\Theta p)$. If an IR regulator is imposed, this momentum cannot vanish and regulates the integral. This can be seen more clearly in the dipole theory (analysed in Section \ref{sec:dip}) where the fixed translation regulates the UV divergence of the integral.

It is important to note that contributions from the $j>1$ terms in Equation \eqref{eqn:def_j_terms} were dropped at the start of this section and are not present in Equation \eqref{eqn:dS_NCYM_result} or elsewhere in these results. However, as is discussed in Appendix \ref{sec:appendix_j_terms}, these do not affect the leading divergence in $S_\textrm{nonplanar}$ or the conclusion that there is no volume law.

In contrast to strong coupling results,
 which saw signs of a volume law for the entanglement entropy even with large regions, this perturbative calculation is only sees an area law. The leading divergence in $S_\textrm{nonplanar}$ is quadratic and proportional to the area of the boundary of the region, $A_\perp$, in line with the area law discussed in the introduction.

\subsection{Charged Scalar}
\label{sec:charged}
For the commutative and the noncommutative theory, the difference when considering a charged scalar comes in at the Wick contraction in Equations \eqref{eqn:comm_wick} and \eqref{eqn:NC_wick} respectively. For the real scalar
\begin{align*}
\lambda \langle \phi(w)\phi(x)\phi(y)\phi(z) \rangle = 
\lambda \left( G_n(w,x) G_n(y,z) 
+ G_n(w,y) G_n(x,z) + G_n(w,z) G_n(x,y) \right),
\end{align*}
whereas for the charged scalar this must be replaced with
\begin{align*}
&\lambda_0 \langle \phi^\dagger(w)\phi(x)\phi^\dagger(y)\phi(z) \rangle
+ \lambda_1 \langle \phi^\dagger(w)\phi(x)\phi(y)\phi^\dagger(z) \rangle \\ 
&= \lambda_0 \left( G_n(w,x) G_n(y,z) + G_n(w,z) G_n(x,y) \right)
+ \lambda_1 \left( G_n(w,x) G_n(z,y) + G_n(w,y) G_n(z,x) \right). 
\end{align*}

In the commutative theory, the fields in the 4-point function are all inserted at the same point, that is $w=x=y=z$. Taking into account the difference in the normalisation of the $\phi^4$ term in the action, the only change is to replace an overall factor of $\frac{3\lambda}{4!}$ by $\frac{2(\lambda_0+\lambda_1)}{4}$. This has no effect on the intermediate steps of the calculation and can just be carried through straight to the final result:
\begin{align*}
S_{1,\textrm{Comm.}} \rightarrow& -\frac{(\lambda_0+\lambda_1) A_\perp \Lambda^2}{3 \cdot 2^{8} \pi^3} 
\ln \frac{\Lambda^2}{4 m^2}.
\end{align*}

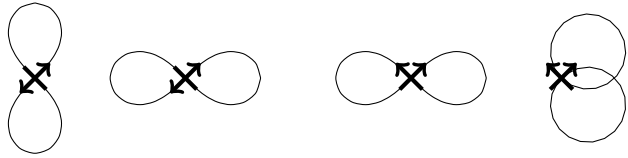
\begin{figure}
  \centering
\begin{tikzpicture}
\draw [ultra thick,<->] (-.2,-.2) -- (.2,.2);
\draw [ultra thick] (-.15,.15) -- (.15,-.15);
\draw [ultra thick,<->,shift={(2,0)}] (-.2,-.2) -- (.2,.2);
\draw [ultra thick,shift={(2,0)}] (-.15,.15) -- (.15,-.15);
\draw [ultra thick,->,shift={(5,0)}] (-.15,-.15) -- (.2,.2);
\draw [ultra thick,<-,shift={(5,0)}] (-.2,.2) -- (.15,-.15);
\draw [ultra thick,->,shift={(7,0)}] (-.15,-.15) -- (.2,.2);
\draw [ultra thick,<-,shift={(7,0)}] (-.2,.2) -- (.15,-.15);
\draw [domain=0:2*pi,smooth,samples=31] plot ( {sin(\x r)*cos(\x r)/(1+sin(\x r)*sin(\x r))} , {cos(\x r)/(1+sin(\x r)*sin(\x r))} );
\draw [domain=0:2*pi,smooth,samples=31] plot ({2+cos(\x r)/(1+sin(\x r)*sin(\x r))} , {sin(\x r)*cos(\x r)/(1+sin(\x r)*sin(\x r))} );
\draw [domain=0:2*pi,smooth,samples=31] plot ( {5+cos(\x r)/(1+sin(\x r)*sin(\x r))} , {sin(\x r)*cos(\x r)/(1+sin(\x r)*sin(\x r))} );
\begin{scope}[shift={(7,0)}]
\draw [domain=0:pi,samples=20,rotate=-45] plot ( {sin(\x r)*cos(\x r)},{sin(\x r) *sin(\x r)});
\draw [domain=0:pi,samples=20,rotate=-45] plot ( {cos(\x r)*cos(\x r)},{cos(\x r) *sin(\x r)});
\end{scope}
\end{tikzpicture}
  \caption[Non-planar diagrams for the charged scalar $\lambda \phi^4$ theory.]{Vacuum bubble diagrams at leading order in the noncommutative charged scalar $\lambda \phi^4$ theory. The two on the left come from the $\lambda_0 \phi^\dagger \star \phi \star \phi^\dagger \star \phi$ term in the action whereas the two on the right from the $\lambda_1 \phi^\dagger \star \phi \star \phi \star \phi^\dagger$ term.}
  \label{fig:charged_scalar}
\end{figure}

For the noncommutative theory, it is a simple matter of writing out the $\star$-products explicitly and following through similar transformations of the integration variables as in the previous section. This procedure gives $2\lambda_0 + \lambda_1$ times the commutative result plus $\lambda_1$ times the result for the nonplanar diagram already encountered for the real scalar. This result can be obtained directly by looking at the 4 diagrams in Figure \ref{fig:charged_scalar} and realising that the only term proportional to $\lambda_1$ can give a nonplanar diagram.

Thus the result for the noncommutative theory with a charged scalar is
\begin{align*}
S_{1,\textrm{NC}} \rightarrow& (2\lambda_0 + \lambda_1) \frac{ A_\perp \Lambda^2}{3 \cdot 2^{9} \pi^3} 
\ln \frac{\Lambda^2}{4 m^2}
-\lambda_1 \frac{ A_\perp \Lambda^2}{3 \cdot 2^{8} \pi^3} 
\frac{-\ln\left(\frac{\Theta^2 m^2 \Lambda^2}{4}\right)}
{1 - \frac{\Theta^2 m^2 \Lambda^2}{4}} 
\end{align*}

\subsection{Dipole theory}
\label{sec:dip}
For the dipole theory, the explicit form of the interaction terms was written out in Equation \eqref{eqn:dip_identities}. Thus,
\begin{align*}
&\ln Z_n = \ln \int \mathcal{D}\phi e^{-S_E[\phi]}\\
&= \ln Z_{n,0} - \int_n d^4 x \left\langle \frac{\lambda_0}4 \Phi^\dagger(x) \Phi(x) \Phi^\dagger(x) \Phi(x) + \frac{\lambda_1}4 \Phi^\dagger(x + \frac12 a) \Phi(x+ \frac12 a) \Phi(x - \frac12 a) \Phi^\dagger(x - \frac12 a) \right\rangle_0 +\ldots 
\end{align*}

Applying Wick's Theorem, using the facts that $G_1(x,x)=G_1(x+a,x+a)$ and $f_n(x+a,x)=f_n(x,x+a)$ (when ignoring the $j>1$ terms) and shifting the integral,
\begin{align*}
\ln Z_{n,1} =& -\frac14 \int_n d^4 x \left[ 2 \lambda_0 G_n(x,x) G_n(x,x)  \right. \\
&+  \left. \lambda_1 \left( 
G_n(x + \frac12 a, x+\frac12 a) G_n(x-\frac12a,x-\frac12a)
+ G_n(x + \frac12 a, x-\frac12 a) G_n(x-\frac12a,x+\frac12a)
\right) \right] \\
S_{1} =& -  \partial_n \left[ \ln Z_{n,1} - n \ln Z_{1,1} \right]_{n=1} \\
=& \frac{2}{4} \partial_n \Bigg( \int d^4 x   
\bigg[ (2 \lambda_0 +\lambda_1) G_1(x,x) n f_n(x,x)  
+ \lambda_1  G_1(x,x+a) n f_n(x,x-a)
\bigg] \Bigg)_{n=1}. 
\end{align*}

Again this is as expected from the diagrammatic approach. Only the single nonplanar diagram gives a new contribution and the 3 planar diagrams give contributions identical to those in the commutative theory.

Focusing on the contribution from the nonplanar diagram, the explicit forms of $G_1$ and $f_n$ give 
\begin{align*}
S_\textrm{nonplanar} = \frac{4 \lambda A_\perp}{12 \pi \cdot 4} \int r dr d\phi \int \frac{d^4 k d p_y d p_z}{(2\pi)^6} \frac{e^{i k_x a}}{k^2 +m^2} 
\int_0^\infty dq q \frac{\partial_\nu [J_{\nu}(qr) J_{\nu}(qr')]_{\nu=0}}{q^2+p_y^2 + p_z^2 +m^2},
\end{align*} 
where now $r'^2 = (\vec{r} - a \hat\imath)^2 = r^2 + a^2 - 2 r a \cos\phi$.

Introducing Schwinger parameters and regulating,
\begin{align*}
S_\textrm{nonplanar} = \frac{\lambda A_\perp}{2^2 3 \pi} \int dr d\phi \frac{d^4 k d p_y d p_z}{(2\pi)^6} dq 
\int_0^\infty d \alpha d \beta 
qr e^{-\alpha k^2 - \beta \left[q^2+p_y^2 + p_z^2 \right] 
- \frac{1}{\alpha \Lambda^2} -\alpha m^2 - \frac{1}{\beta \Lambda^2}  - \beta m^2} \\
e^{i k_x a} \partial_\nu [J_{\nu}(qr) J_{\nu}(qr')]_{\nu=0}.
\end{align*} 
In this case, all the momenta integrals except $q$ are Gaussian,
\begin{align*}
S_\textrm{nonplanar} =& \frac{\lambda A_\perp}{2^{8} 3 \pi^4} \int dr d\phi dq d\alpha d\beta 
\frac{qr}{\alpha ^2 \beta} e^{-\frac{a^2}{4 \alpha} 
- \frac{1}{\alpha \Lambda^2}  -\alpha m^2
- \beta q^2 - \frac{1}{\beta \Lambda^2} -\beta m^2 }
\partial_\nu [J_{\nu}(qr) J_{\nu}(qr')]_{\nu=0}.
\end{align*} 

The $\alpha$ integral can be factored out to give, using Equation \eqref{eqn:ET_146-25},
\begin{align*}
\int_0^\infty d \alpha \frac{e^{-\frac{1}{\alpha} \left(\frac{a^2}{4} + \frac{1}{\Lambda^2} \right) -\alpha m^2 }}{\alpha^2} 
= \int_0^\infty d \alpha e^{-\alpha \left(\frac{a^2}{4} + \frac{1}{\Lambda^2} \right) 
-\frac{m^2}{\alpha} }
=& \frac{2m}{\sqrt{\frac{a^2}{4} + \frac{1}{\Lambda^2}}} 
K_1 \left( 2 m \sqrt{\frac{a^2}{4} + \frac{1}{\Lambda^2}} \right) \\
\xrightarrow{\Lambda \rightarrow \infty}& 
\frac{4m}{a} 
K_1 \left( m a \right) \\
\xrightarrow{m \rightarrow 0}& 
\frac{4}{a^2} 
\end{align*}
This factor came from evaluating $G_1(0,a \hat\imath)$ which goes as $\sim \frac{1}{a^2}$ as expected. The fixed nonlocality scale has regulated the UV divergence in this case. In the dipole theory the distance of the translation is fixed, as opposed to the non-commutative case where the translation is proportional to the momentum in the $y$-direction which can vanish in the IR. 

Using Equation \eqref{eqn:DLMF_10.22.E67},
\begin{align*}
S_\textrm{nonplanar} =& -\frac{\lambda A_\perp }{2^{7} 3 \pi^4 a^2} \int_0^\infty d\beta 
\left[ \int_0^\infty dr \int_0^{2 \pi} d\phi  
\frac{r}{\beta^2} e^{ - \frac{r^2+r'^2}{4\beta} }
K_0 \left(\frac{r r'}{2\beta} \right) \right]
e^{- \frac{1}{\beta \Lambda^2} -\beta m^2}.
\end{align*}

Rescaling $r \rightarrow a r$ and $\beta \rightarrow a^2 \beta$ to make them dimensionless ($r' \rightarrow a\sqrt{r^2 + 1 -2r\cos\phi}$ under this) and  defining $H(\beta)$ as the part of the previous equation enclosed in brackets,
\begin{align*}
S_\textrm{nonplanar}=& -\frac{\lambda A_\perp }{2^{7} 3 \pi^4 a^2} \int_0^\infty d\beta 
H(\beta)
e^{- \frac{1}{\beta a^2 \Lambda^2} -\beta a^2 m^2}.
\end{align*} 

$H(\beta)$ is dimensionless and finite for $\beta \in (0,\infty)$. The integrand is exponentially suppressed for small $\beta$ and numerical evaluation of the $r$ and $\phi$ integrals confirm that $H(\beta) \xrightarrow{\beta \rightarrow 0} 0$. The other potential source of a divergence is at large $\beta$ and numerical integration finds that $H(\beta) \xrightarrow{\beta \rightarrow \infty} \frac{2\pi}{\beta}$ leading to a logarithmic divergence at large $\beta$ that must be regulated by $e^{-\beta a^2m^2}$,
\begin{align*}
\int^\infty \frac{d\beta}{\beta} e^{-\beta a^2m^2} = -\ln(a^2m^2) + O(m^0),
\end{align*}
to leading order in the small $m$ limit.

Thus $S_\textrm{nonplanar}$ has only an IR divergence in the dipole theory. The leading divergence in the $j=1$ term is
\begin{align*}
S_\textrm{nonplanar} = -\frac{\lambda A_\perp }{3 \cdot 2^{6} \pi^3 a^2} 
\left[ -\ln(a^2 m^2) \right],
\eqnlabel{eqn:dS_dip_result}
\end{align*}
however there will be contributions to this order from the $j>1$ terms which were dropped. The  the conclusion of this analysis is that the nonplanar diagram does not contribute to the leading divergence of entanglement entropy at this order as it is subleading to the contribution from the planar diagram.

The nonlocality introduced in the dipole theory does not affect the area law, as the total entanglement entropy at this order in perturbation theory is dominated by the planar diagrams which matched the result from the commutative theory. Even the subleading terms we have analysed do not follow any sort of volume law as they are not proportional to the lengthscale of the nonlocality.
The only effect of the nonlocality is to regulate the UV divergence otherwise present. Similar behaviour was observed in \cite{Minwalla:1999px}, where one of the ways that the nonlocality manifested itself was by softening divergences in nonplanar diagrams.

\section{Final remarks}
\label{sec:conclusion}
In this paper we computed the first perturbative correction to the entanglement entropy in two nonlocal theories, a $\phi^4$ theory defined on the noncommutative plane and a dipole theory. 

The contribution to the entanglement entropy in each of these theories at first order in coupling comes from vacuum bubble diagrams. The planar diagrams give the same contribution in all three theories. However, the nonplanar diagram is affected by the modified $\star$-product. Never the less, these diagrams do not modify the area observed in the commutative theory. Thus, at this order in perturbation theory and for the region considered at least, all these theories follow an area law with no sign of a volume law, as opposed to the strongly coupled case where the signature of the volume law could be seen even for large regions.

In the commutative theory it has been shown that the modification to the entanglement entropy at first order in perturbation theory can be absorbed into the renormalisation of the mass \cite{Hertzberg:2012mn}. It would be interesting to see if a similar interpretation can be made in the case of the theories considered here.

Finally, a comment about the commutative limit. Since the quantities dealt with in the paper are not UV finite, this is not a straightforward issue. The general pattern is that the nonlocality has served as an additional regulator that softens certain divergences. Thus, if the nonlocality is removed, these divergences reappear and the commutative limit applied to the final results is not smooth. 


\section*{Acknowledgements}
I am grateful to my supervisor Joanna Karczmarek for her support and advice. I am also grateful to Philippe Sabella Garnier, Mark Van Raamsdonk and Keshav Dasgupta for useful discussions. I was supported during this work by the Natural Sciences and Engineering Research Council of Canada's Alexander Graham Bell Canada Graduate Scholarship.

\appendix
\section{Analysis of the potential divergences from the $j>1$ terms}
\label{sec:appendix_j_terms}
This analysis follows that of \cite{Calabrese:2004eu}, where it is found that the leading divergence when the Green's function is evaluated at coincident points is entirely contained in the $j=1$ term.

The Green's function for the scalar field on the $n$-sheeted space was given in Equation \eqref{eqn:def_j_terms}:
\begin{align*}
G_n(x,x') = &\int_0^\infty \frac{dk}{\pi}\int \frac{d^{d_\perp} p^\perp}{(2 \pi)^{d_\perp}} 
\int_0^\infty dq q \frac{J_{k}(qr) J_{k}(qr')}{q^2+p^2_\perp+m^2} \cos(k (\theta-\theta')) 
e^{i p_\perp (x_\perp - x'_\perp)} \\
&-\frac1{12\pi n^2}\int \frac{d^{d_\perp} p^\perp}{(2 \pi)^{d_\perp}} 
\int_0^\infty dq q \frac{\partial_\nu [J_{\nu}(qr) J_{\nu}(qr')]_{\nu=0}}{q^2+p^2_\perp+m^2} 
e^{i p_\perp (x_\perp - x'_\perp)} \\
&-\sum_{j>1} \frac{B_{2j}}{\pi n^{2j} (2j)!}\int \frac{d^{d_\perp} p^\perp}{(2 \pi)^{d_\perp}} 
\int_0^\infty dq q \frac{(\partial_\nu)^{2j-1} 
[J_{\nu}(qr) J_{\nu}(qr') \cos(\nu(\theta-\theta'))]_{\nu=0}}{q^2+p^2_\perp+m^2} 
e^{i p_\perp (x_\perp - x'_\perp)}. 
\end{align*}
The first term is independent of $n$ and did not enter into the calculation of the entanglement entropy. The second term was the subject of our investigation. However, the third term was dropped with the claim that it could not introduce any new divergences. This appendix will justify this claim.

We start by revisiting the entanglement entropy in the commutative theory. In this case from Equation \eqref{eqn:S1_comm} 
\begin{align*}
S \sim \int r dr G_1(r,r) f_n(r,r),
\end{align*}
where only the contributions to the divergences in the final result have been kept.

The Green's function when evaluated at coincident points gives a $\Lambda^2$ divergence
\begin{align*}
G_1(r,r) \sim& \int d^4 p \frac{1}{p^2+m^2} 
\sim \int d\alpha ~ p^{3} dp ~ e^{-\alpha(p^2+m^2) - \frac{1}{\alpha \Lambda^2}} \\
\sim& \Lambda^2 - m^2 \log\Lambda^2.
\end{align*}

The $f_n$ term has the form
\begin{align*}
 f_n(r,r) \sim& \int d^{2} p^\perp ~ q dq ~ \frac{\partial_\nu [J_{\nu}(qr) J_{\nu}(qr)]_{\nu=0}}{q^2+p^2_\perp+m^2} 
+ \sum_{j>1} \int d^{2} p^\perp ~ q dq ~ \frac{\partial_\nu^{2j-1} [J_{\nu}(qr) J_{\nu}(qr)]_{\nu=0}}{q^2+p^2_\perp+m^2}.
\end{align*}
The momentum integrals can be evaluated when the function is evaluated at coincident points
\begin{align*}
\int & d^{2} p^\perp ~ q dq ~ \frac{ J_{\nu}(qr) J_{\nu}(qr)}{q^2+p^2_\perp+m^2} 
= \int d\beta ~ p dp ~ q dq ~ J_{\nu}(qr) J_{\nu}(qr) e^{-\beta (q^2+p^2+m^2) -\frac{1}{\beta \Lambda^2}} \\
&\sim e^{-\frac12 r^2} I_\nu(\frac12 r^2) \log\Lambda^2.
\end{align*}
This must be integrated over $r$
\begin{align*}
\int_0^\infty rdr e^{-\frac12 r^2} I_\nu(\frac{1-\epsilon^2}{2} r^2) 
=  \frac{1}{\sqrt{2} \epsilon}  - \nu +O(\epsilon),
\end{align*}
where a small $\epsilon$ has been added to regulate the integral. It is only divergent because $\partial_\nu^{2j-1}$ was passed though the integral sign. Once this derivative is applied, $\epsilon$ can be safely taken to zero. A calculation of terms $O(\Lambda^0)$ would require a more careful analysis, but this is sufficient for extracting the leading $O(\log \Lambda^2)$ divergence. Thus
\begin{align*}
\int d^4 x f_n(x,x) \sim A_{\perp} \log\Lambda^2 \left[ \partial_\nu (-\nu)  + \sum_{j>1}  \partial_\nu^{2j-1} (-\nu) \right] = A_{\perp} \log\Lambda^2 \left[ -1  + \sum_{j>1}  0 \right].
\end{align*}

This shows that all the $j>1$ terms vanish when the Green's function is evaluated at coincident points and the divergence is entirely contained in the $j=1$ term.

In the noncommutative and dipole theories, the Green's functions are evaluated at points separated by the length scale of the nonlocality rather than at coincident points. However, we saw that the source of divergences in the entanglement entropy was regions in the integral where this separation vanishes. This analysis shows that these divergences are contained in the $j=1$ term.


\bibliographystyle{JHEP}
\bibliography{../refs}{}

\end{document}